\documentclass[
    a4paper,
    12pt,
    english,
    headings=standardclasses]
{scrartcl}

\usepackage{tabularx}
\usepackage{orcidlink}
\usepackage{authblk}
\usepackage{lineno}
\usepackage[utf8]{inputenc}
\usepackage[T1]{fontenc}
\usepackage{graphicx}
\usepackage{amsmath,amssymb,amsfonts}
\usepackage{amsthm}
\usepackage{mathrsfs}
\usepackage[title]{appendix}
\usepackage{xcolor}
\usepackage{textcomp}
\usepackage{booktabs}
\usepackage{listings}
\usepackage{nameref}
\usepackage{caption}
\usepackage{hyperref}
\hypersetup{colorlinks=true, linkcolor=blue, citecolor=blue, urlcolor=blue, filecolor=black}
\usepackage[capitalise]{cleveref}
\usepackage[backend=biber,
style=nature,
hyperref=true,
]{biblatex}
\bibliography{DeptDyn}

\AtEveryBibitem{
    \ifentrytype{online}{}{
        \clearfield{urlyear}
        \clearfield{urlmonth}
        \clearfield{urlday}
    }
}

\DeclareMathOperator*{\argmin}{arg\,min}
\DeclareMathOperator{\wks}{wKS}
\DeclareMathOperator{\SLL}{SLL}
\DeclareMathOperator{\SLN}{SLN}

\begin{document}

\newcommand{\papertitle}{How large should academic departments be?}
\title{\Large\papertitle}

\author[1,2$*$]{Jan Bachmann\orcidlink{0000-0002-6153-4714}}

\author[2]{Lisette Esp\'in-Noboa\orcidlink{0000-0002-3945-2966}}
\author[1,2]{Fariba Karimi\orcidlink{0000-0002-0037-2475}}
\author[3,4,5]{Daniel B. Larremore\orcidlink{0000-0001-5273-5234}}
\author[6,7$\dagger$]{Gerardo I\~niguez\orcidlink{0000-0001-7181-5520}}
\author[3,4,5$\dagger$]{Aaron Clauset\orcidlink{0000-0002-3529-8746}}

\affil[1]{Graz University of Technology, 8010 Graz, Austria}
\affil[2]{Complexity Science Hub, 1030 Vienna, Austria}
\affil[3]{Department of Computer Science, University of Colorado Boulder, Boulder, CO, USA}
\affil[4]{BioFrontiers Institute, University of Colorado Boulder, Boulder, CO, USA}
\affil[5]{Santa Fe Institute, Santa Fe, NM, USA}
\affil[6]{Tampere Complexity Lab, Data Science Research Centre, Tampere University, 33720 Tampere, Finland}
\affil[7]{Centro de Ciencias de la Complejidad, Universidad Nacional Aut\'onoma de M\'exico, 04510 Ciudad de M\'exico, Mexico}

\affil[$*$]{\small{\textit{Corresponding author email: jan.bachmann@tugraz.at}}}
\affil[$\dagger$]{\small{\textit{These authors contributed equally to this work.}}}

\newcommand{\starsection}[1]{\section*{#1}\setcurrentname{#1}\addcontentsline{toc}{section}{#1}}
\newcommand{\starsubsection}[1]{\subsection*{#1}\setcurrentname{#1}\addcontentsline{toc}{subsection}{#1}}

\newcommand{\NDepts}{14,418}
\newcommand{\NFac}{294,084}
\newcommand{\NInst}{419}
\newcommand{\SizeQMed}{12}
\newcommand{\SizeGTFifty}{1.3\%}
\newcommand{\NBS}{1,000}
\newcommand{\ShareExtremes}{12\%}
\newcommand{\NDeptExtremes}{1,700}
\newcommand{\NDeptGTFifty}{180}
\newcommand{\NDeptCentral}{10,000}
\newcommand{\ShareDeptROne}{5.6\%}
\newcommand{\ClosureRiskRegOne}{8\%}
\newcommand{\ClosureRiskRegThree}{2\%}
\newcommand{\NClosureRegOne}{588}
\newcommand{\NExposureRegOne}{6,770}
\newcommand{\NSurviveRegOne}{6,182}
\newcommand{\GrowthRegOneMax}{1.4}

\newcommand{\NFacRaw}{360,248}
\newcommand{\NDeptRaw}{17,111}
\newcommand{\NInstRaw}{439}
\newcommand{\NField}{112}
\newcommand{\NMultiFac}{8,229}
\newcommand{\NMultiApp}{38,385}
\newcommand{\NNonDept}{2,137}
\newcommand{\NBreakDept}{50}

\setkomafont{section}{\normalfont\large\bfseries}
\setkomafont{subsection}{\normalfont\bfseries}
\setkomafont{subsubsection}{\normalfont\bfseries}

\renewcommand{\Authfont}{\fontsize{12}{14.4}\selectfont}

\makeatletter
\newcommand\setcurrentname[1]{\def\@currentlabelname{#1}}
\makeatother

\date{\vskip -2em}

\captionsetup{format=plain,parindent=0pt}

\flushbottom
\maketitle

\begin{abstract}
    Academic departments are the primary unit of scholarship and education at universities, and they vary vastly in their sizes.
    However, the consequences and natural dynamics of department size are poorly understood.
    Small departments face disproportionate teaching and administrative overhead per faculty member, while large ones face coordination costs and thematic incoherence.
    Here, we characterize and model the dynamics of academic department sizes using $14,000$ U.S.-based departments in eight academic domains.
    Across all domains, similar broad-tailed distributions reveal a common size range from 4 to 23 faculty members, widening across domains at its upper border.
    Annual size-dependent closure risks and growth rates indicate that stability is greatest in this size range: below it, small departments either close or grow quickly; within it, closure risk is low and sizes stabilize; above it, large departments can persist, with marginal attrition and minimal closure risk.
    An analytically tractable model of size-dependent coagulation and fragmentation, informed only by the aggregated size distribution, reproduces department dynamics across the full size range.
    Rescaling each domain by its most stable size reveals a common regression toward the stable range across most domains.
    Our results establish academic departments as organizations with natural size dynamics defined by a grow-or-close pattern for the smallest departments, and a weak pressure against unlimited growth for the largest.
\end{abstract}

\pagebreak

\starsection{Introduction}
\label{sec:intro}
In 2023, West Virginia University faced protests from both its students and faculty for its decision to cut roughly 8\% of its academic offerings, including about 140 faculty positions~\cite{spitalniak_wvuboardapproves_2023}.
This cut led to the closure of its Department of World Languages, Literatures and Linguistics, then home to 24 faculty, after a review flagged it for low and declining student demand~\cite{quinn_westvirginiasunprecedented_2023,wvutech_wvutechacademic_}.
University restructuring is not always so dramatic, and department closures can and do occur at all department sizes in different forms.
For instance, in an effort to pool resources into a larger unit, the University of Texas at San Antonio merged its Departments of Demography and Sociology in 2023~\cite{amandacerreto_mergerutsasdemography_2023}, while Cornell University closed its Department of Computational Biology by splitting it into two departments with greater growth potential, in response to rising domain demand~\cite{melanielefkowitz_newdatascience_2018}.

Restructuring academic departments by closures, mergers, and splits can stem from external pressures and internal adaptations, and all such changes are ultimately reflected in department size.
For instance, external pressure from governments, funding agencies, and the public~\cite{gumport.sporn_institutionaladaptationdemands_1999,perkins_organizationfunctionsuniversity_1972,edwards_academicdepartmenthow_1999,walvoord.etal_academicdepartmentshow_2000} can suppress growth by demanding greater efficiencies, or it can support growth by providing resources.
Competitions around student enrollment, faculty hiring, and research funding~\cite{aagaard.etal_concentrationdispersalresearch_2020,walvoord.etal_academicdepartmentshow_2000} could further grow already successful departments.
More broadly, shifts in societal and academic trends can alter demand and support across entire disciplines, e.g., the decades-long growth of STEM fields over the humanities, and affect the availability of talent and financing~\cite{hur.etal_recenttrendsus_2017,smith.etal_graduateenrollmentpostdoctoral_2025,gowder_usefulstatstrends_2024,burke_scienceengineeringindicators_2019}, and hence department sizes.
As the ``embodiment of a discipline at the university''~\cite{edwards_academicdepartmenthow_1999}, departments directly face these pressures while also having to accommodate domain-specific equipment and teaching expectations~\cite{walvoord.etal_academicdepartmentshow_2000}.

Size is a fundamental variable for organizational dynamics~\cite{josefy.etal_allthingsgreat_2015} and while the size of a department constrains its structure and outcomes, including research performance~\cite{leitner.etal_impactsizespecialisation_2007}, its determinants and consequences are poorly understood.
Gibrat's law of proportionate growth treats size changes as size-independent, yielding right-skewed size distributions from noise alone without stabilizing to any particular size~\cite{gibrat_inegaliteseconomiques_1931}.
Departments adapt their size through hiring and attrition, structured by domain-specific prestige hierarchies~\cite{clauset.etal_systematicinequalityhierarchy_2015,wapman.etal_quantifyinghierarchydynamics_2022,fitzgerald.etal_temporaldynamicsfaculty_2023}.
As hiring opportunities open with retirements, funding, and student demand, department growth may fluctuate independently of size.
Certain extensions of Gibrat's model add a lower size boundary that keeps the smallest organizations from shrinking further, or a threshold below which they close~\cite{gabaix_zipfslawgrowth_1999,luttmer_selectiongrowthsize_2007}, enabling a narrow size-dependence for closure and growth at small sizes.
Real department closures, however, are not confined to the smallest sizes, as illustrated by the case of the department split at Cornell~\cite{melanielefkowitz_newdatascience_2018}.
Closure at a moderate-sized department at Cornell was a coordinated effort to facilitate faculty growth and support more specialized student interest.

Unlike Gibrat's law, organizational ecology theory treats the empirical size distribution as the outcome of selection pressures that act differently on small and large organizations~\cite{hannan.freeman_structuralinertiaorganizational_1984}.
Small departments may bear disproportionate administrative overhead per member~\cite{brinkman.leslie_economiesscalehigher_1986,tirivayi.etal_sizeeconomiesscale_2014,edwards_academicdepartmenthow_1999} and face resource scarcity~\cite{baker.cullen_administrativereorganizationconfigurational_1993}, while still staffing a minimum range of courses.
Under this `liability of smallness'~\cite{barron.etal_timegrowtime_1994,aldrich.auster_evendwarfsstarted_1986}, they must either grow quickly toward a more stable size or eventually close.
Small size may itself make consolidation easier, as the few faculty and courses involved can be redistributed with relatively less disruption to other university operations.
In contrast, large departments may enjoy `economies of scale', as shared infrastructure and administration lower the overhead cost per member and scaling large courses requires few additional faculty~\cite{josefy.etal_allthingsgreat_2015}, along with the resources and prestige to attract funding.
This may imply an underlying tendency for departments to grow continually, to reap the benefits of scale.
Yet costs related to coordination, bureaucratic complexity~\cite{blau_formaltheorydifferentiation_1970} and thematic incoherence~\cite{baker.cullen_administrativereorganizationconfigurational_1993} may also accumulate with greater size and eventually limit continued growth.
Large departments may further resist restructuring of any kind, an organizational inertia that grows with size~\cite{hannan.freeman_structuralinertiaorganizational_1984} and, through their reliability, lowers their risk of closure.
Together, these opposing pressures predict growth and closure rates that vary across the size spectrum, pulling departments toward a stable range of sizes where the forces roughly balance.
The particular location or range of this stable region may differ by domain, due to discipline-specific differences in required equipment, available funding, and student demand.
Whether departments exhibit a stable range of sizes can be established from examining how their sizes change over time within the overall size distribution.
A stable range would indicate clearly whether department sizes are governed by the size-dependent selection of organizational ecology or by the size-independent growth of Gibrat's law.

Here, we characterize the size distributions and dynamics of over 14,000 departments across eight academic domains at U.S. PhD-granting universities between 2011 and 2023 to evaluate whether Gibrat's law or organizational ecology better explains observed department sizes.
The size distribution across all departments and within each domain is well captured by two similar, parsimonious distributions with a single peak, stretching into a long tail of large departments.
Year-to-year size changes reveal that closure risks and growth rates depend on size across the entire size spectrum, departing from Gibrat's law and its boundary extensions and pulling departments toward a stable range, consistent with organizational ecology.
Small departments below the range either close or grow quickly, departments within the stable range stagnate, and large departments above it persist under weak net attrition.
We emulate the fitted distributions with a simple coagulation--fragmentation model in which departments grow and shrink by the flow of single faculty members at size-dependent rates~\cite{becker.doring_kinetischebehandlungkeimbildung_1935,wattis_introductionmathematicalmodels_2006}.
The model's size-dependent rates recover the direction and turning points of the empirical size changes, stabilizing large departments while varying in magnitude across domains at small sizes.
Departments regulate their size, through hiring and attrition; mergers and splits are the visible extremes of that process.

\starsection{Results}
\label{sec:results}
We address how pressures towards growth and closure shape academic department sizes by analyzing \NDepts~departments across eight academic domains at~\NInst~PhD-granting universities based in the United States.
These data are derived from a census of tenured and tenure-track faculty collected by the Academic Analytics Research Center.
After removing secondary affiliations, non-academic job titles, and non-department organizations, our sample covers the department affiliations of~\NFac~tenured and tenure-track faculty members between 2011 and 2023 (see SI Note \nameref{ssec:data_processing}).
We sort departments into the following eight domains based on a prior categorization of these same data~\cite{wapman.etal_quantifyinghierarchydynamics_2022}: Humanities, Applied Sciences, Education, Medicine and Health, Social Sciences, Engineering, Natural Sciences, and Mathematics and Computing (see~\Cref{tab:domain_summary} for sample sizes and~\Cref{stab:domain_categorization} for a domain-field mapping).
We then obtain department sizes $s_t$ as the count of affiliated faculty members in year $t$.
To aggregate at the department level, we calculate the static size distribution $p_s$ over each department's median size across the observation period.

\begin{figure}[ht!]
    \centering
    \includegraphics[width=\textwidth]{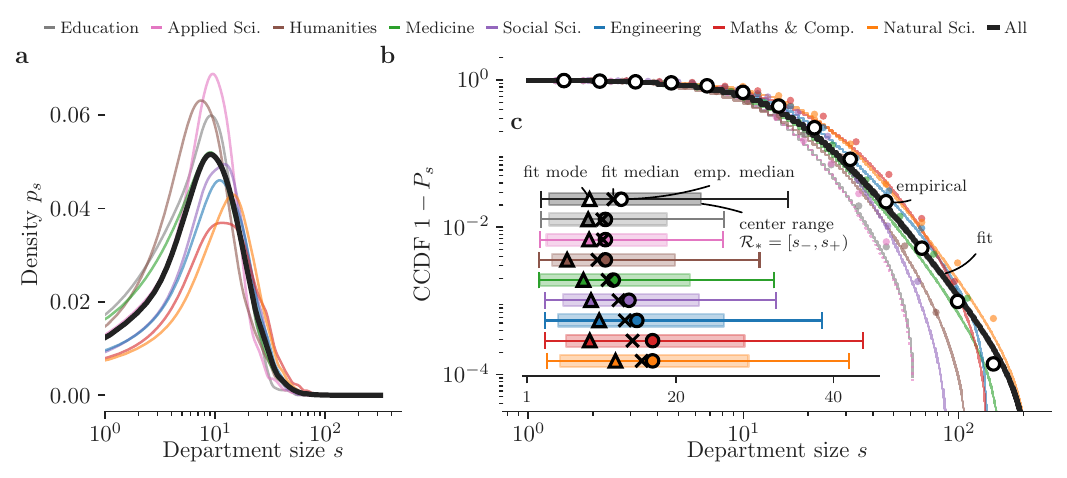}
    \caption{\small \textbf{Fitting department size distributions across domains.} (\textbf{a}) Empirical size distribution $p_s$, for all departments and per domain.
    Both small and large departments are common, with few above 100 tenured and tenure-track faculty.
    The resulting right skew places the median size of~\SizeQMed~above the peak at around 10 faculty.
    Domains share similar tails but differ in the prevalence of small and intermediate sizes.
    (\textbf{b}) The three-parameter shifted log logistic ($\SLL$) captures these features and plausibly describes the empirical distribution ($\wks \approx 0.02$ at $p \approx 0.08$; see~\nameref{sec:methods}).
    Only Humanities, Social Sciences, and Mathematics and Computing are better described by shifted log normals ($\SLN$) at equal parameter count, whose faster decaying tail limits extreme sizes.
    (\textbf{c}) Fitted modes and central ranges per domain, asymmetric and varying in width (whiskers at $90\%$-quantiles; shaded band is the central range; see~\nameref{sec:methods}).
    Modes lie at the lower end of these ranges, suggesting dynamics that pull smaller departments toward the mode while leaving larger ones less affected.
    }\label{fig:figure1}
\end{figure}

\starsubsection{Evidence of common department size distributions across domains}
\label{ssec:size_distribution}
Across the U.S. academic system, the size distribution $p_s$ shows a single peak at around ten faculty members and falls off toward both extremes (\Cref{fig:figure1}.\textbf{a}).
The fall-off, however, is asymmetric: small sizes form a plateau, while large sizes stretch into a pronounced long tail.
Around~\NDeptExtremes~departments (\ShareExtremes) have only four or fewer tenured or tenure-track professors affiliated, while the largest~\ShareExtremes~host 26 or more faculty.
Within this group of large departments, about~\NDeptGTFifty~departments contain more than 50 faculty.
The largest departments also tend to be hosted by the largest universities (Spearman's $r_s = 0.50;\ p < 0.001$; see~\Cref{stab:inst_dept_size}).
Even a department of around ten faculty is responsible for a substantial number of students.
Linking each department to its institution's undergraduate enrollment, we estimate that each faculty member is responsible, on average, for roughly 16--28 undergraduate students (the interquartile range across institutions; \Cref{stab:student_enrollment}).
Hence, a department at the peak of around ten faculty then is responsible for roughly 160--280 undergraduate majors.
This load varies strongly across institutions and decreases only mildly with department size.

In each of the eight domains, the same peaked shape recurs, but variably shifted and stretched along the size axis (\Cref{fig:figure1}.\textbf{a}).
These shifts are systematic, as the size distributions differ significantly between all pairs of domains (Benjamini-Hochberg corrected~\cite{benjamini.hochberg_controllingfalsediscovery_1995} permutation tests on the tail-weighted Kolmogorov-Smirnov distance $\wks$, $p < 0.05$; see~\nameref{sec:methods};~\cite{clauset.erwin_evolutiondistributionspecies_2008}).
The Humanities, Education, Applied Sciences, and Medicine and Health have median sizes below the overall median, while the Social Sciences, Engineering, Mathematics and Computing, and Natural Sciences fall above it (see also~\Cref{stab:size_descriptives} for an overview).
The differences between domains concentrate at large sizes and are more pronounced in domains that depend on expensive infrastructure (e.g., Engineering and Natural Sciences).
For instance, departments at the $95\%$-quantile size for Mathematics and Computing are around $1.7$ times larger than those in Education.
Small departments are equally common across domains, but large departments are more prevalent in some domains than others, shifting their median sizes upward (\Cref{stab:size_descriptives}).

The shared shape across domains suggests a single generative process whose parameters vary by domain.
To narrow down its form and to quantify each domain's mode and central range, we fit ten candidate distributions drawn from the literature on firm sizes and organizational ecology (see~\Cref{stab:models} for a distribution overview;~\cite{kleiber.kotz_statisticalsizedistributions_2003}).
We accept a fit only if the empirical distribution could plausibly be drawn from the fitted distribution (see~\nameref{sec:methods} for the fitting and goodness-of-fit test~\cite{clauset.etal_powerlawdistributionsempirical_2009}).
Of the ten considered, two distributions capture the empirical sizes across all departments and most domains: the shifted log logistic ($\SLL$) and the shifted log normal ($\SLN$), whose fitted density we denote as $q_s$.
Both require only three parameters (shape, scale, and a location shift), where the shift matches the plateau observed among the smallest departments (\Cref{fig:figure1}.\textbf{b}; see~\Cref{sfig:model_fit_heatmap} for a fit overview and~\Cref{stab:models} for their functional form).
The log logistic fits slightly better across most domains, although the two are difficult to separate statistically~\cite{diyali.etal_discriminatinglognormalloglogistic_2024}:\
they differ mainly in their tails, where the log logistic decays as a power law and the log normal falls off faster, assigning lower probability to the existence of very large departments.

For most department sizes, both distributions coincide and imply similar modes and central ranges for all domains (\Cref{fig:figure1}.\textbf{c} and~\Cref{tab:domain_summary}; see~\nameref{sec:methods} for further details).
The fitted mode marks the peak of the distribution, a candidate size that departments may be pulled toward.
We test this hypothesis against the dynamics below.
Modes lie in the range of 8--12 faculty members across most domains, with Humanities the only outlier at about six.
The central range $\mathcal{R}_* = [s_-, s_+)$, measured as one geometric standard deviation around the median, runs from a lower boundary $s_- \approx 4$ to an upper boundary $s_+ \approx 23$ and holds roughly~\NDeptCentral~($72\%$) of all departments.
Because both distributions have multiplicative spread, the central range extends farther above the median than below it, with the mode near its lower end and consistently below both the empirical and fitted medians.
Most departments are larger than the mode.
This asymmetry suggests that size-related pressures constrain small departments more strongly than large ones.
Across domains, the lower boundary $s_-$ differs by only a few faculty members in absolute terms, from three in Medicine and Health, to six in Mathematics and Computing, while the upper boundary $s_+$ varies more strongly with domain-specific equipment, funding, and student demand, commonly reaching 29 faculty members in Mathematics and Computing.

\begin{table}[t]
\centering
\caption{\small \textbf{Statistics of departments at the domain level.}
For each domain we report the number of departments ($N$) and represented universities ($N_\text{univ}$), the fitted distribution ($\SLL$ or $\SLN$), the fitted mode $s^*$, and the central range ($\mathcal{R}_*$) defined as one geometric standard deviation around the fitted median.
}
\label{tab:domain_summary}
\begin{tabularx}{\linewidth}{Xrrrrr}
\toprule
Domain & $N$ & $N_\mathrm{univ}$ & Fit & Mode $s^*$ & Central range $\mathcal{R}_*$ \\
\midrule
All & 14,418 & 419 & $\mathrm{SLL}$ & 9.0 & $[3.8,\,23.1)$ \\
Education & 816 & 238 & $\mathrm{SLL}$ & 8.8 & $[3.9,\,18.8)$ \\
Applied Sciences$^\dagger$ & 1,646 & 254 & $\mathrm{SLL}$ & 8.9 & $[3.5,\,18.8)$ \\
Humanities & 2,962 & 310 & $\mathrm{SLN}$ & 6.1 & $[4.3,\,19.9)$ \\
Medicine and Health & 2,368 & 302 & $\mathrm{SLL}$ & 8.2 & $[2.7,\,21.7)$ \\
Social Sciences & 1,905 & 298 & $\mathrm{SLN}$ & 9.2 & $[5.7,\,22.9)$ \\
Engineering & 1,337 & 269 & $\mathrm{SLL}$ & 10.2 & $[5.0,\,26.1)$ \\
Mathematics and Computing & 1,123 & 340 & $\mathrm{SLN}$ & 9.0 & $[6.0,\,28.7)$ \\
Natural Sciences & 2,970 & 379 & $\mathrm{SLL}$ & 12.3 & $[5.3,\,29.2)$ \\
\bottomrule
\multicolumn{5}{l}{$^\dagger$ -- Empirical distribution not a plausible draw from fit ($p< 0.05$)}\\
\end{tabularx}

\end{table}

\starsubsection{Size-dependent closure and growth do not follow proportionate growth}
\label{ssec:size_dynamics}
The size distribution alone does not reveal the mechanism, as a single-peaked, right-skewed shape can arise from size-dependent regulation as well as from proportionate growth and its boundary extensions~\cite{mitzenmacher_briefhistorygenerative_2004}.
To separate them, we characterize how departments close or grow over time, testing whether their sizes drift toward the fitted mode or independently of size as Gibrat's law predicts.
A department closure marks a major restructuring of that department, e.g., through an outright shutdown, a merger into another unit, or a split into several units.
To capture all such events, we count a closure whenever a department's size drops from $s>0$ in year $t$ to $s=0$ faculty in year $t+1$.
We then estimate the closure risk $c_s$ as the share of departments at size $s$ that close (see~\nameref{sec:methods}).
In contrast, department growth reflects the net of regular hiring and attrition, provided a department does not close.
We measure a department's growth rate $g_s$ as the ratio of its sizes in consecutive years $g_s = s_{t+1}/(G_t \, s_{t})$, accounting for the system-wide growth $G_t$.
What distinguishes the candidate mechanisms is whether closure risk $c_s$ or growth rate $g_s$ depends on size.

\begin{figure}[t]
    \centering
    \includegraphics[width=\textwidth]{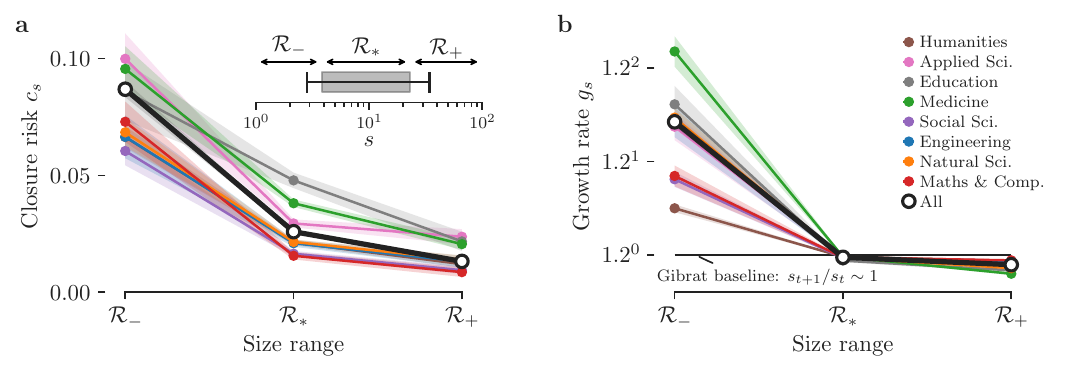}
    \caption{\small \textbf{Regions of department size change.} Annual department size changes $s_t \rightarrow s_{t+1}$ arise from hiring, attrition, and closure.
    (\textbf{a}) We sort departments into three ranges by size, labeled small ($\mathcal{R}_-$), central ($\mathcal{R}_*$), and large ($\mathcal{R}_+$) on the axis; the central range spans one geometric standard deviation around the median (inset; see~\nameref{sec:methods}).
    For a department of size $s_t$, we record its closure if $s_{t+1} = 0$ and estimate the conditional risk $c_{s}$ as the closure count over all years and departments at size $s_t = s$.
    Error bands show one standard error around the per-range mean.
    As departments grow bigger, their closure risk reduces for all domains, with the largest survival gain occurring from the small to the central range ($\mathcal{R}_- \to \mathcal{R}_*$).
    Small departments face a closure risk of roughly~\ClosureRiskRegOne, reducing to below~\ClosureRiskRegThree~for large departments above it.
    (\textbf{b}) If small departments survive ($s_{t+1} > 0$), they tend to grow ($g_s > 1$).
    On average, surviving small Medicine and Health departments grow \GrowthRegOneMax-fold from one year to the next.
    Growth slows down within the central range and transitions to marginal net attrition above it for all domains.
    With their size-dependence defying Gibrat's law~\cite{gibrat_inegaliteseconomiques_1931}, the empirical closure and growth together pull departments towards the central range, making it stable.
    Large departments remain stable while small ones either close or grow.}
    \label{fig:figure2}
\end{figure}

Gibrat's law of proportionate growth~\cite{gibrat_inegaliteseconomiques_1931} treats annual size changes as a multiplicative and size-independent process, $s_{t+1} = \eta_t\, s_t$, with the growth factor $\eta_t$ drawn independently from the current size $s_t$.
Under this assumption, log normal distributions emerge from multiplicative noise.
Boundary extensions of Gibrat's model retain proportionate growth but add a small-size boundary or closure threshold~\cite{gabaix_zipfslawgrowth_1999,luttmer_selectiongrowthsize_2007}, raising closure risk at small sizes and using proportionate growth elsewhere.
These hypotheses differ in how the growth ratio $g_s$ varies with size.
Gibrat's proportionate growth predicts $g_s = 1$ for all $s$.
Its small-size boundary extensions keep $g_s = 1$ above a small boundary size, deviating only below it.
In contrast, organizational ecology and economies of scale would instead predict size-dependent pressures $g_s > 1$ below the central range and $g_s < 1$ above it, pulling departments toward that range.
We test these predictions across three ranges: the central range $\mathcal{R}_*$ and the ranges below ($\mathcal{R}_-$) and above ($\mathcal{R}_+$) it, split at its boundaries $s_-$ and $s_+$ (see~\nameref{sec:methods} and inset of~\Cref{fig:figure2}.\textbf{a}).

Small departments are markedly less stable than large ones, facing a substantially higher annual risk of closure.
Departments below the central range face an annual closure risk of roughly~\ClosureRiskRegOne, compared to less than~\ClosureRiskRegThree~for those above it (\Cref{fig:figure2}.\textbf{a}).
Out of~\NExposureRegOne~annual observations of departments in the small range, fully \NClosureRegOne~of them close.
Closure risks declining with department size could arise from independent per-faculty attrition alone, as a department of size $s_t$ requires $s_t$ simultaneous departures to close, i.e. reach $s_{t+1} = 0$, and smaller departments require fewer departures.
This explanation assumes that every faculty member leaves independently at a common annual attrition risk.
The risk does not vary with department size, so a department of size $s$ closes only when all $s$ of them leave in the same year.
Recovered from the closure risk as $\sqrt[s]{c_s}$, the per-faculty risk would then be constant across sizes.
Instead, we find it to grow with $s$, reaching implausibly high values for large departments (\Cref{sfig:closure_risk_expectation}).
Together with the consistently declining closure risk with increasing department size ranges, department closure instead appears as a structural phenomenon that disproportionately affects small departments.
Closure risk declining with department size is consistent with organizational ecology, where the reliability of larger, established departments lowers their risk of closure~\cite{hannan.freeman_structuralinertiaorganizational_1984}.

Conditional on survival, small departments tend to grow, while departments at intermediate sizes stabilize and large ones shrink slightly (\Cref{fig:figure2}.\textbf{b}).
We measure department growth as the size ratio $g_s = s_{t+1} / (G_t \, s_{t})$ for all transitions from size $s_t$ to (non-zero) size at year $t+1$, de-trended by the system-wide annual growth $G_t$, binned by sorting $s_t$ into the size ranges, and averaged $g_s$ across the size range bins (see~\nameref{sec:methods}).
Out of the same~\NExposureRegOne~small-range observations, the~\NSurviveRegOne~ departments that avoid closure grow 1.3-fold on average.
Growth levels decrease within the central range and turn into marginal net attrition above it, with error bands excluding neutral growth in both outer ranges (\Cref{fig:figure2}.\textbf{b}).
The growth rates are similar when not conditioning on survival (see~\Cref{sfig:growth_rates}).

The size-dependent closure and growth we observe in the empirical data are inconsistent with both Gibrat's rule of proportionate growth, which predicts no such size dependence, and the small-size boundary extensions of Gibrat's model, which predict deviations only among the smallest department sizes.
Instead, the dynamics follow the predictions of organizational ecology, where growth and closure tend to pull departments toward the central range and weakly hold them there.
Although the central range $\mathcal{R}_*$ has been fitted from the static distribution alone, its low closure risk and balanced growth make it dynamically stable, so the geometric and the dynamically stable range coincide.

We refer to $\mathcal{R}_*$ as the stable range of department sizes.
Below it, elevated closure risk combines with fast conditional growth, as we would expect from what the liability of smallness predicts for small organizations~\cite{barron.etal_timegrowtime_1994,aldrich.auster_evendwarfsstarted_1986}.
Within it, closure risks reduce sharply and growth settles, as expected by the benefits of economies of scale~\cite{josefy.etal_allthingsgreat_2015}.
Above it, minimal closure risks and only marginal net attrition leave large departments to persist under a weak downward pull, consistent with the structural inertia expected to grow with size~\cite{hannan.freeman_structuralinertiaorganizational_1984}.
The measured size-dependent regulation thus concentrates below the stable range and largely fades above it, confirming the asymmetry that the static distributions $p_s$ (and their fits $q_s$) suggested.

\begin{figure}[ht!]
    \centering
    \includegraphics[width=\textwidth]{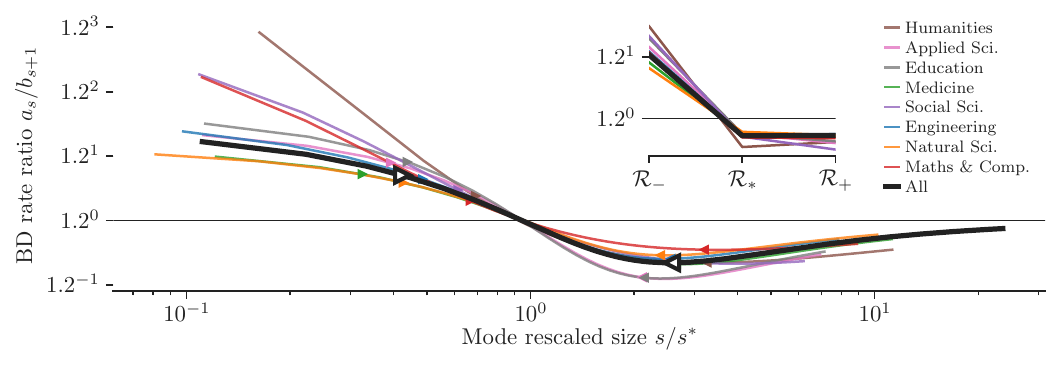}
    \caption{\small \textbf{Modeling department size dynamics with coagulation/fragmentation mechanisms.}
    In the Becker-Döring (BD) model, clusters (departments) of size $s$ grow and shrink by gaining or losing single particles (faculty members) at size-dependent rates $a_{s}$ and $b_{s}$, respectively~\cite{becker.doring_kinetischebehandlungkeimbildung_1935}.
    At its equilibrium, the net flow between department counts of consecutive sizes $s$ and $s+1$ is zero.
    Under this assumption, the fitted distributions directly determine the rate ratio $a_s / b_{s+1}$, which indicates net growth above one and net attrition below one (see also~\nameref{sec:methods}).
    While derived from the static fits alone, the BD rates match the direction of the measured dynamics (inset aggregates the x-axis by the size range of~\Cref{fig:figure2}).
    For departments below the stable range $\mathcal{R}_*$, growth dominates; through the stable range, fragmentation takes over at the stable size $s^*$.
    The net fragmentation surplus across the stable range mirrors the combined intermediate closure risks and stagnating growth.
    From the upper edge of the stable range onward, both rates approach balance, leaving only a small fragmentation surplus at large sizes.
    Departments this big face only mild net attrition.}
    \label{fig:figure3}
\end{figure}

\starsubsection{A coagulation--fragmentation model recovers the empirical size dynamics}
\label{ssec:becker_doering}
To interpret these dynamics mechanistically, we treat the fitted $\SLL$ and $\SLN$ distributions as equilibria of a Becker–D\"oring model~\cite{becker.doring_kinetischebehandlungkeimbildung_1935}, a classical framework for size-dependent coagulation--fragmentation dynamics.
In this model, a department of size $s$ can either gain a single faculty member to grow to size $s+1$ (coagulation) or lose one faculty member to shrink to size $s-1$ (fragmentation), at size-dependent rates $a_s$ and $b_s$,
\begin{equation}
    \label{eq:bd_harpoons}
    D_s + D_1 \rightleftharpoons^{a_s}_{b_{s+1}} D_{s+1} \enspace,
\end{equation}
where $D_1$ is a single faculty member drawn from a common reservoir.
Assuming the fitted densities $q_s$ represent the model's equilibrium, the flow of departments into and out of each size must balance.
Their consecutive size ratio then directly determines the ratio of coagulation to fragmentation rates,
\begin{equation}
    \label{eq:bd_ratio}
    \frac{a_s}{b_{s+1}} = \frac{q_{s+1}}{q_s}
\end{equation}
(\Cref{fig:figure3}; see~\nameref{sec:methods}).
We normalize this ratio so that coagulation and fragmentation balance ($a_s = b_{s+1}$) at the fitted mode $s^*$.

The size-dependent balance of coagulation and fragmentation that is predicted on the basis of the model's static fits nicely reproduces the empirically measured dynamics across all three ranges (\Cref{fig:figure3}).
At equilibrium, the rate ratio $a_s/b_{s+1}$ is the model counterpart of the empirical growth ratio $g_s$ and closure risk $c_s$, above one where coagulation dominates and below one where fragmentation does.
It starts above one for the smallest departments, where growth is strongest and closure risk highest, crosses one at the fitted mode $s^*$ as growth slows and closure risk falls, and dips to a minimum across the stable range before slowly re-approaching one from below at the largest sizes, where closure risk is minimal.
The three ranges follow this shape, with a coagulation surplus below the stable range ($\mathcal{R}_-$) and only marginal net attrition above it, under which large departments can persist ($\mathcal{R}_+$).

When rescaled by the fitted mode $s^*$, the coagulation-to-fragmentation curves overlap across the stable range and above it, while diverging across domains below it (\Cref{fig:figure3}).
With the exception of Applied Sciences and Education, domains differ mainly in their fitted mode $s^*$.
Below the stable size, coagulation exceeds fragmentation by a factor ranging from about $1.2$ to $1.7$ at the smallest sizes, indicating that the liability of smallness varies in strength across domains.
At the largest sizes, the ratio approaches one from below, so that size-dependent regulation concentrates at small and intermediate sizes and decays toward the tail.
This alignment reinforces what the common shape of $p_s$ across domains had already suggested: a single process governs department size dynamics, with domains differing mainly in their mode $s^*$ and in the strength of coagulation at the smallest sizes.

\starsection{Discussion}
\label{sec:discussion}
Human organizations are widely observed to exhibit right-skewed size distributions~\cite{gibrat_inegaliteseconomiques_1931,kleiber.kotz_statisticalsizedistributions_2003}.
We show that this empirical pattern also holds for academic departments (\Cref{fig:figure1}).
However, fitting models to static distributions alone does not provide sufficient evidence to identify their generative mechanisms, as the same skew can arise from various size dynamics.
These include size-independent variations and proportionate growth, as predicted by Gibrat's law, or size-dependent regulation, as predicted by organizational ecology~\cite{mitzenmacher_briefhistorygenerative_2004,gabaix_zipfslawgrowth_1999,luttmer_selectiongrowthsize_2007}.
Detailed data on annual department size dynamics reveal that department growth and closure clearly depends on department size across the entire size spectrum (\Cref{fig:figure2}).
This size dependence defies Gibrat's proportionate growth model and instead supports the size-dependent selection of organizational ecology~\cite{hannan.freeman_structuralinertiaorganizational_1984}, which pulls departments toward a stable range of sizes.
Across domains, this regulation takes a common form.

Closure risk is highest among the smallest departments, falling considerably as departments become larger.
Moreover, how risk falls with department size follows the same size-dependent pattern in every domain (\Cref{fig:figure2}.\textbf{a} and~\Cref{sfig:closure_risk_expectation}).
Academic domains instead differ mainly in the baseline level of this risk, which correlates with their size scale.
Education, Applied Sciences, and Medicine \& Health have the lowest small-size stable-range boundaries $s_-$ (\Cref{tab:domain_summary}), placing their small departments among the most closure-prone.
While closure risks similarly decline with department size for all domains, growth rates differ mainly in how strongly small departments grow.
Within the stable range of department sizes, growth balances department sizes and nets into a mild tendency to decline at larger sizes (\Cref{fig:figure2}.\textbf{b}).
The coagulation--fragmentation model reproduces these patterns (\Cref{fig:figure3}).
When rescaled by the fitted mode $s^*$, its rate ratios overlap across and above the stable range, separating only below it.
A single size-dependent model therefore explains the department size dynamics across the U.S. academic system.

Below the stable range, small departments exhibit both elevated growth rates and higher closure risks (\Cref{fig:figure2}), consistent with the `liability of smallness' described in organizational ecology~\cite{barron.etal_timegrowtime_1994,aldrich.auster_evendwarfsstarted_1986}.
These liabilities may include disproportionate administrative overhead per faculty member~\cite{gumport.sporn_institutionaladaptationdemands_1999,perkins_organizationfunctionsuniversity_1972,edwards_academicdepartmenthow_1999,brinkman.leslie_economiesscalehigher_1986,tirivayi.etal_sizeeconomiesscale_2014} and the fixed cost of staffing a minimum curriculum~\cite{walvoord.etal_academicdepartmentshow_2000}.
Beyond these costs, small departments may lack the settled routines and institutional standing that give larger, established units their reliability, leaving them both more exposed to closure and less disruptive to dissolve~\cite{hannan.freeman_structuralinertiaorganizational_1984,aldrich.auster_evendwarfsstarted_1986}.
Small departments also teach slightly more undergraduates, on average, per faculty member than larger departments, implying a larger mentoring load per faculty (\Cref{stab:student_enrollment}).

The liabilities of smallness also seem to vary across domains, as both the closure liability and growth rates at small sizes are larger in some domains than in others.
For instance, Applied Sciences, Medicine \& Health, and Education programs are more often oriented toward professional training in
fields like business, nursing, and teaching.
This difference may drive larger (market-driven) fluctuations in student demand and a greater incentive by universities to reconfigure department names and rosters more fluidly in order to follow those trends, making these programs
smaller and more closure-prone.
In contrast, Engineering and Natural Sciences programs are often more oriented toward educating students in established academic fields.
This focus may mitigate fluctuations in student demand and lower the incentive for universities to reconfigure programs and faculty rosters away from the traditional field categories, making these departments larger and less prone to closure.
Moreover, domain variation in closure coincides with the fastest small-department growth in Education and Medicine \& Health, which may reflect a greater flow of faculty between practice and academia, making their small departments grow when faculty flow from practice to academia, and dissolve when the flows largely reverse.
Other domains, in contrast, may draw faculty mainly from a slower doctoral pipeline, where the boundary between practice and academia is less porous.
Identifying which of these plausible mechanisms drives the difference would require additional data not available in this study.

Unlike the case of small-size dynamics, the stabilization of medium-sized and large departments takes the same form in every domain.
We find intermediate size ranges to be dynamically stable, as the size-dependent effects of closure risk and growth rates balance each other at these size ranges (\Cref{fig:figure2,fig:figure3}).
At these intermediate sizes, departments appear to be largely free of the liabilities of smallness, and large enough to balance the benefits of economies of scale against the costs of coordination~\cite{josefy.etal_allthingsgreat_2015}.
In academia, benefits of scale can arise from shared infrastructure and administration, and from teaching efficiencies, where course sizes can be larger and faculty can specialize in specific parts of the curriculum.
Above the stable range, coordination, bureaucratic complexity, and thematic incoherence can induce rising social, epistemic, and curricular costs that together induce a pressure to shrink~\cite{baker.cullen_administrativereorganizationconfigurational_1993,blau_formaltheorydifferentiation_1970}, even as there may yet be other bureaucratic efficiencies that come from larger size.
Still, large departments have the lowest closure risk (\Cref{fig:figure2}.\textbf{a}).
Organizational ecology predicts that their settled routines, such as established programs, both lower their closure risk and resist change~\cite{hannan.freeman_structuralinertiaorganizational_1984}.
The mild net attrition of larger departments (\Cref{fig:figure2}.\textbf{b}) may reflect this structural inertia, combined with the diseconomies of coordination and incoherence that grow with size.

Our findings come with several limitations in how we measure and model department size and its dynamics.
First, we measure department size as only the count of tenured and tenure-track faculty at U.S.\ PhD-granting universities, which may only be a correlate of the quantities that directly determine departmental stability, such as student demand or available budget.
Beyond the institution-level enrollment considered here (\Cref{stab:student_enrollment}), records linking departments to enrollment and funding would show whether size itself drives the measured dynamics or whether it just proxies resource constraints.
Second, the Becker--D\"oring model treats the observed size distribution as stationary and identifies only the ratio of coagulation (merging) to fragmentation (splitting), leaving open how each rate depends on size on its own.
Long-term data on faculty turnover would estimate the two separately, reveal whether the balance at intermediate and large sizes reflects the low turnover expected under structural inertia~\cite{hannan.freeman_structuralinertiaorganizational_1984}, and clarify the stationarity of the department size distribution.
Third, the model's single-member exchanges also do not represent department closures, mergers, and splits.
Measuring closure risk by size-zero transitions also counts these events together with renames, potentially inflating the reported closure risk.
Records of explicit department restructuring would better distinguish closures, mergers, splits and renames, and would reveal how restructuring may vary by department size.

Our results position academic departments as evolving organizations whose sizes are shaped by different organizational forces, depending on their size.
These sizes are not determined by the proportionate, size-independent growth of Gibrat's law.
Instead, department sizes follow the size-dependent dynamics predicted by organizational ecology, with the strongest effects falling on the smallest departments.
Because many departments sit close to the closure-prone small size range, they are likely to experience organizational ecology pressures.
Affected departments typically follow a clear pattern: they either grow quickly to a stable size, or they close.
Large departments, in contrast, experience a looser set of pressures, with only a weak tendency to regress toward typical department sizes.
The equilibrium of a simple coagulation--fragmentation model condenses these dynamics across most domains into one common, size-balancing process operating at domain-specific scales (\Cref{fig:figure3}).
In connecting organizational ecology with higher education research, these findings provide a foundation for future work on organizational adaptation in academia.
\starsection{Materials and methods}
\label{sec:methods}

\starsubsection{Fitting department size distributions}
From the departments and faculty remaining after filtering (see SI Note~\nameref{ssec:data_processing}), we define department size $s_t$ as the number of tenured and tenure-track faculty affiliated with a department in year $t$.
We characterize each department by its typical size $\tilde{s}$, defined as the median of $s_t$ across all observed years, and form the empirical size distribution $p_s = N_s / N$, where $N_s$ is the number of departments with $\tilde{s} = s$ and $N = \sum_s N_s$.

To identify which parametric family best describes $p_s$, we fit a set of standard size distributions~\cite{kleiber.kotz_statisticalsizedistributions_2003} to the data.
For each candidate distribution $m$ with probability mass function $q_{s,m}$, we compute the tail-weighted Kolmogorov--Smirnov statistic $\wks$ between the empirical distribution $p_s$ and $q_{s,m}$
\begin{equation}
    \wks_m = \max_s \frac{|P_s - Q_{s,m}|}{\sqrt{P_s(1-P_s)}},
\end{equation}
with $P_s$ and $Q_{s,m}$ being the cumulative distribution functions of $p_s$ and $q_{s,m}$, respectively.
By emphasizing the noisy tail, $\wks$ imposes a stricter goodness-of-fit threshold than the unweighted KS statistic~\cite{clauset.erwin_evolutiondistributionspecies_2008,press.etal_numericalrecipesart_1992}.
We find the best fitting parameters by minimizing $\wks$ over distribution parameters $\theta_m$ as
\begin{equation}
    \theta^*_m = \argmin_{\theta} \wks(p_s, q_{s,m}(\theta)),
\end{equation}
using differential evolution, a global optimization algorithm~\cite{storn.price_differentialevolutionsimple_1997} with a tolerance level of $10^{-8}$ and a maximum of 10,000 iterations.
Parameter bounds are set to ensure valid distribution parameter values (see~\Cref{stab:models} for a distribution overview).

The optimized $\wks_m$ value alone does not indicate whether $q_{s,m}$ is a plausible distribution for the observed $p_s$.
To assess the goodness of fit without relying on parametric assumptions, we test whether the observed $\wks$ is within expectation if $p_s$ were an empirical sample drawn from the fitted $q_{s,m}$.
We compare the observed $\wks$ to a null distribution of values obtained by generating synthetic size counts from the fitted distribution and refitting the respective distribution for each synthetic dataset.
This way, we assess the expected $\wks$ values if the data were generated by the fitted distribution~\cite{clauset.etal_powerlawdistributionsempirical_2009}.
We generate~\NBS~synthetic samples $\hat n_{s} \sim \mathrm{Multinomial}[q_{s,m}(\theta^*_m)]$ of size $N$ for each fitted distribution $m$ with optimized parameters $\theta^*_m$ and compute $\hat p_s = \hat n_{s} / N$.
We then fit the candidate distribution to each synthetic sample to obtain $\hat \theta^*_m$ and the corresponding $\widehat{\wks}$ value, constructing the null distribution of $\wks$ under the fitted distribution.
A $p$-value is finally computed as the fraction of $\widehat{\wks}$ values that are greater than or equal to the observed $\wks$.

Although goodness of fit varies considerably across domains, shifted variants of log normal ($\SLN$) and log logistic ($\SLL$) distributions balance good fits (lowest $\wks$ values and $p > 0.05$) with the simplicity of only three parameters (shape, scale, and shift), in contrast to more complex distributions such as the shifted Burr (see~\Cref{sfig:model_fit_heatmap} for a comparison of fitted $\wks$ and $p$ values across candidate distributions and~\Cref{stab:fit_params} for the fitted parameters of the $\SLL$ and $\SLN$ distributions across domains).

\starsubsection{Comparing size distributions across domains}
\label{ssec:domain_comparison}
To test whether size distributions differ between domains, we run a permutation test on the $\wks$ distance for each pair of domains.
For each pair, we pool the median sizes of both domains' departments, randomly reassign them to two synthetic domains of the original domain sizes, and recompute the $\wks$ distance between the two resulting distributions.
The $p$-value is the fraction of 1,000 permuted distances that reach or exceed the observed one.
We correct the $p$-values for multiple testing across the domain pairs with the Benjamini--Hochberg procedure~\cite{benjamini.hochberg_controllingfalsediscovery_1995}.

\starsubsection{Distribution characteristics}
\label{ssec:dist_characteristics}
To characterize the typical department size and its variation under each fitted distribution, we translate the fitted parameters from the log axis back to the real size axis~$s$.
In both $\SLL$ and $\SLN$, the log-transformed shifted size $\log(s + s_0)$ follows a symmetric distribution: a Logistic distribution for $\SLL$ and a Normal distribution for $\SLN$,
\begin{equation}
\label{eq:sll_sln}
    \log(s + s_0) \sim \mathrm{Logistic}(\log \alpha,\, 1/\beta) \quad \text{and} \quad \log(s + s_0) \sim \mathrm{Normal}(\mu,\, \sigma).
\end{equation}
Because these distributions are symmetric on the log axis, their location parameters ($\log\alpha$ and $\mu$) give the median of $\log(s+s_0)$, and their standard deviations on the log axis are
\begin{equation}
    \sigma_{\log}^{\SLL} = \frac{\pi}{\beta \sqrt{3}} \quad \text{and} \quad \sigma_{\log}^{\SLN} = \sigma.
\end{equation}

To express the central range on the real size axis $s$, we exponentiate $\sigma_{\log}$.
The additive log-axis standard deviation then becomes a multiplicative factor $\sigma_{\mathrm{geo}} = \exp(\sigma_{\log})$ around the median of $s + s_0$, which is $\alpha$ for $\SLL$ and $e^\mu$ for $\SLN$.
To express the range in real department sizes $s$, we finally subtract the shift $s_0$ to compute the median, mode $s^*$, and central range $\mathcal{R}_* = [s_-, s_+)$ on the real size axis as summarized in~\Cref{tab:dist_characteristics}.
Note that the $\SLL$ mode requires $\beta > 1$, which is satisfied for all distribution fits (see~\Cref{stab:fit_params}).

\begin{table}[h!]
    \centering
    \caption{\label{tab:dist_characteristics} Median, mode, and central range for the $\SLL$ and $\SLN$ distributions in terms of their fitted parameters.}
    \begin{tabularx}{\linewidth}{Xccc}
        \toprule
        Distribution & Median & Mode $s^*$ & Range $\mathcal{R}_*$\\
        \midrule
        $\SLL$ & $\alpha - s_0$ & $\alpha \Bigl(\frac{\beta - 1}{\beta + 1}\Bigr)^{1/\beta} - s_0$ & $\left[\alpha / \sigma_{\mathrm{geo}}^{\SLL} - s_0,\; \alpha\,\sigma_{\mathrm{geo}}^{\SLL} - s_0\right)$\\
        $\SLN$ & $e^\mu - s_0$ & $e^{\mu - \sigma^2} - s_0$ & $\left[e^\mu / \sigma_{\mathrm{geo}}^{\SLN} - s_0,\; e^\mu\,\sigma_{\mathrm{geo}}^{\SLN} - s_0\right)$\\
        \bottomrule
    \end{tabularx}
\end{table}

\starsubsection{Closure risk and growth rates}
\label{ssec:size_growth_closure}
For every department size $s_t$ at year $t < T$ prior to the last observation year $T = 2023$, we analyze department size dynamics from closure risks and growth rates.
We estimate the closure risk $c_s := \Pr(s_{t+1} = 0 \mid s_t = s)$ as the share of observations at size $s_t = s$ that transition to size zero by year $t+1$.
Annual growth rates are measured by $g_s := s_{t+1} / (G_t \, s_t)$ over yearly department sizes with $s_t = s$ and $s_{t+1} > 0$.
Here, $G_t = S_{t+1} / S_t$, with $S_t$ as the total number of faculty across all departments in year $t$, is the global system growth ratio from year $t$ to $t+1$ used to de-trend each department's growth.

We group closure risks and growth rates into three size ranges based on the central range from the fitted distribution: small departments $\mathcal{R}_- = [1,\, s_-)$, mid-sized departments $\mathcal{R}_* = [s_-,\, s_+)$, and large departments $\mathcal{R}_+ = [s_+,\, \infty)$.
For each size range we report the closure risk's average and binomial standard error, and the growth rate's geometric mean and geometric standard error.
We require at least 50 observations per range across all years to ensure reliable estimates.

\starsubsection{Becker--D\"oring dynamics}
To link static size distributions and observed dynamics, we consider the Becker--D\"oring (BD) model of size-dependent coagulation and fragmentation processes~\cite{becker.doring_kinetischebehandlungkeimbildung_1935}, in which size-$s$ departments $D_s$ merge with a size-one faculty unit $D_1$ to grow or shed a size-one unit to shrink.
Its size dynamics are described by the master equation for the number of departments of size $s$ at time $t$, denoted $n_s(t)$,
\begin{equation}
    \frac{dn_s}{dt} = a_{s-1} n_{s-1} n_1 - a_s n_s n_1 + b_{s+1} n_{s+1} - b_s n_s\ .
\end{equation}
Here $a_s$ is the coagulation rate at which a size-$s$ department gains a member and grows to size $s+1$, and $b_s$ is the fragmentation rate at which a size-$s$ department sheds a member and shrinks to size $s-1$.

At equilibrium, coagulation and fragmentation balance such that $a_s\, n_s\, n_1 = b_{s+1}\, n_{s+1}$, yielding the rate ratio $a_s / b_{s+1} = n_{s+1} / (n_s\, n_1)$.
To compute this ratio from data, we assume the observed size distribution $p_s$ to be in equilibrium and substitute the fitted distribution $q_s$ for the concentrations, setting $n_s = q_s / q_1$.
This reduces the rate ratio to
\begin{equation}
    \frac{a_s}{b_{s+1}} = \frac{q_{s+1}}{q_s}\ .    
\end{equation}
Choosing $n_s = q_s / q_1$ sets only the absolute scale of $a_s/b_{s+1}$, and through it the location of the crossover point at the mode; the $s$-dependence (shape) is unaffected.

Note that the equilibrium condition identifies only the rate ratio.
Multiplying both $a_s$ and $b_s$ by the same constant leaves the stationary distribution unchanged.
The empirical annual growth rates are also not directly comparable to these rates, because they exclude departments that closed during the year and they aggregate multiple individual hiring and attrition events into a single annual change.
Comparisons between the rate ratio and the empirical growth rates are therefore limited to the direction of net change at each size, not its magnitude.

\starsection{Data and code availability}
\label{sec:data_avail}
An aggregate data record containing annual department sizes as faculty counts is available from~\url{https://doi.org/10.5281/zenodo.21456430}.
From this record, the code in~\url{https://github.com/mannbach/dept_size_dynamics} reproduces the results.

\starsection{Author contributions}
\label{sec:author_contrib}

\noindent
Jan Bachmann conceived the study, curated the data, performed the formal analysis, led the investigation, developed the methodology and software, supervised the project, validated the results, produced the visualizations, and wrote the original draft. Daniel~B.~Larremore contributed to conceptualization, data curation, and formal analysis. Gerardo I\~niguez contributed to formal analysis, methodology, and supervision. Aaron Clauset contributed to conceptualization, formal analysis, methodology, supervision, and visualization. Fariba Karimi contributed to conceptualization, interpretation, and funding acquisition. Lisette Esp\'in-Noboa contributed to the conceptualization and interpretation of the results, and presented them at a conference. All authors reviewed and edited the manuscript.

\starsection{Acknowledgments}
\label{sec:acknowledgements}
We thank Chethan Kavaraganahalli Prasanna for his support in processing the data, Jula L\"uhring for her feedback on the manuscript, Sam Zhang for providing data to link the AARC data with student enrollment counts, and Bernhard Geiger for his feedback on the distribution fitting.

Jan Bachmann was partially supported by ERC Starting grant NetFair no.~101165497.

\starsection{References}
\printbibliography[heading=none]

\clearpage
\setcounter{section}{0}
\setcounter{figure}{0}
\setcounter{table}{0}
\renewcommand{\thesection}{S\arabic{section}}
\renewcommand{\thefigure}{S\arabic{figure}}
\renewcommand{\thetable}{S\arabic{table}}

\thispagestyle{plain}
\vspace*{4cm}
\begin{center}
    {\Large\itshape Supplementary Information for}\\[2ex]
    {\Large\bfseries \papertitle}
\end{center}
\clearpage

\starsection{Data processing}
\label{ssec:data_processing}
The Academic Analytics Research Center (AARC) provides an annual census of faculty affiliations at U.S. PhD-granting universities.
Each record links a faculty member to a department in a given year and carries the member's academic rank, appointment type, and the department's field.
The sample holds \NFacRaw~individuals across \NDeptRaw~departments at \NInstRaw~institutions over the 13 years from 2011 to 2023.
Fields are organized in a hierarchy of \NField~fields nested within 26 areas and eight domains, following a prior categorization of U.S. faculty hiring~\cite{wapman.etal_quantifyinghierarchydynamics_2022} (see~\Cref{stab:domain_categorization} for the domain-field mapping).

We restrict the sample to tenured and tenure-track faculty in academic departments through three filters at the level of individual appointments.
First, we keep only records whose rank is Assistant Professor, Associate Professor, or Professor, excluding non-tenure-track titles.
Second, we keep primary appointments only and drop secondary affiliations, so that each faculty member contributes to a single department per year.
This removes \NMultiApp~appointments held by the \NMultiFac~faculty members who appear in more than one department.
Third, we drop organizational units other than departments, identified by their name matching any of School, College, Center, Institute, or Program, which removes \NNonDept~units.

Finally, we require each department to be observed without gaps in its faculty coverage across the years it appears, removing \NBreakDept~departments whose department-year records break.
We re-derive the appointment and attrition events after this step and iterate until no further departments are affected.
The processed sample covers \NFac~tenured and tenure-track faculty affiliated with \NDepts~departments at \NInst~institutions between 2011 and 2023.
From these records we compute annual department sizes $s_t$ and their median-based aggregate distribution as described in the main text.

\begin{table}[ht!]
\centering
\caption[Domain categorization]{\label{stab:domain_categorization}
\textbf{Domain categorization.}
For each domain we report the assigned fields based on a prior categorization~\cite{wapman.etal_quantifyinghierarchydynamics_2022}. Some departments may be categorized into multiple areas.}
\begin{tabularx}{\linewidth}{Xlr}
\toprule
Domain & Area & $N$ \\
\midrule
Natural Sciences & Biological Sciences & 1,626 \\
 & Chemical Sciences & 519 \\
 & Earth Sciences & 465 \\
 & Physical Sciences & 440 \\
 & Unclassified & 128 \\
Humanities & Language, Literature, Culture & 1,127 \\
 & History & 295 \\
 & Philosophy & 278 \\
 & Theology and Religion & 229 \\
 & Arts & 153 \\
 & Linguistics & 88 \\
 & Unclassified & 872 \\
Medicine and Health & Health & 1,229 \\
 & Medical Sciences & 806 \\
 & Unclassified & 366 \\
Social Sciences & Psychological Sciences & 387 \\
 & Sociology & 378 \\
 & Economics & 317 \\
 & Political Science & 308 \\
 & Anthropology & 233 \\
 & Geography & 159 \\
 & Gender Studies & 92 \\
 & Unclassified & 116 \\
Applied Sciences & Business & 971 \\
 & Agriculture & 273 \\
 & Architecture, Design, Planning & 116 \\
 & Unclassified & 286 \\
Engineering & Engineering & 1,140 \\
 & Unclassified & 197 \\
Mathematics and Computing & Computational Sciences & 666 \\
 & Mathematical Sciences & 451 \\
 & Unclassified & 41 \\
Education & Education & 577 \\
 & Unclassified & 239 \\
\bottomrule
\end{tabularx}

\end{table}

\clearpage
\begin{table}[h]
\centering
\begin{tabularx}{\linewidth}{Xrrrr}
\toprule
Domain & $N$ & $Q_{5}$ & Median & $Q_{95}$ \\
\midrule
Natural Sciences & 2,970 & 4.0 & 16.0 & 41.0 \\
Humanities & 2,962 & 2.0 & 10.0 & 32.0 \\
Medicine and Health & 2,368 & 2.0 & 11.0 & 33.0 \\
Social Sciences & 1,905 & 3.0 & 13.0 & 32.0 \\
Applied Sciences & 1,646 & 2.0 & 11.0 & 27.0 \\
Engineering & 1,337 & 3.0 & 13.0 & 39.1 \\
Mathematics and Computing & 1,123 & 3.5 & 15.0 & 43.0 \\
Education & 816 & 3.0 & 11.0 & 25.0 \\
All & 14,418 & 3.0 & 12.0 & 34.0 \\
\bottomrule
\end{tabularx}

\caption[Descriptive size statistics]{\label{stab:size_descriptives}
\textbf{Descriptive size statistics.}
For each domain, and for all departments pooled, we report the number of departments $N$, the 5\% and 95\% percentiles ($Q_{5}$ and $Q_{95}$) and median size across departments' time aggregated median sizes $\tilde s$.}
\end{table}

\begin{table}[h]
\centering
\begin{tabularx}{\linewidth}{Xrrrr}
\toprule
Total faculty at university & $N$ & $Q_{25}$ & Median & $Q_{75}$ \\
\midrule
$[1, 25)$ & 38 & 4.5 & 6.5 & 9.4 \\
$[25, 125)$ & 76 & 6.2 & 9.0 & 13.0 \\
$[125, 625)$ & 166 & 9.0 & 10.0 & 13.5 \\
$\geq 625$ & 139 & 12.0 & 14.0 & 16.0 \\
All & 419 & 8.9 & 11.0 & 14.5 \\
\bottomrule
\end{tabularx}

\caption[Institution versus department size]{\label{stab:inst_dept_size}
\textbf{Institution versus department size.}
We group universities by their total faculty size and report the median and inter-quartile range over all median department sizes $\tilde s$ hosted by the respective universities. Both sizes correlate (Spearman $r_s = 0.495,\ p = 2.4 \times 10^{-27}$; Pearson $r_p = 0.200,\ p = 3.8 \times 10^{-5}$).}
\end{table}

\begin{table}[h]
\centering
\begin{tabularx}{\linewidth}{Xlrrrr}
\toprule
Department size & Range & $N$ & $Q_{25}$ & Median & $Q_{75}$ \\
\midrule
Small $\mathcal{R}_-$ & $[1, 4)$ & 1,036 & 15.9 & 21.2 & 28.6 \\
Stable $\mathcal{R}_*$ & $[4, 24)$ & 10,850 & 15.9 & 20.3 & 28.2 \\
Large $\mathcal{R}_+$ & $\geq 24$ & 2,050 & 13.2 & 18.6 & 24.2 \\
All &  & 13,936 & 15.5 & 20.2 & 27.5 \\
\bottomrule
\end{tabularx}

\caption[Student enrollment]{\label{stab:student_enrollment}
\textbf{Student enrollment.}
We use the U.S. Department of Education's College Scorecard data~\cite{u.s.departmentofeducation_collegescorecarddata_2026}, which provides annual undergraduate enrollment counts per institution, to estimate an undergraduate-student-to-faculty ratio and attribute it to our departments. We link 400 of our~\NInst~institutions by matching institution names; of these, 367 report an undergraduate degree-seeking headcount. For an institution $i$, let $\hat u_i$ be its median undergraduate enrollment over time and $F_i$ its total faculty, the sum of its departments' median sizes $\hat s$. We assign every department the ratio $\hat u_i / F_i$ of its institution and, for each department-size range, report the median and IQR of this ratio over the departments in it. Each institution contributes once per department it has in the range.}
\end{table}

\clearpage
\starsection{Distribution fitting}
\begin{figure}[ht]
    \centering
    \includegraphics[width=\textwidth]{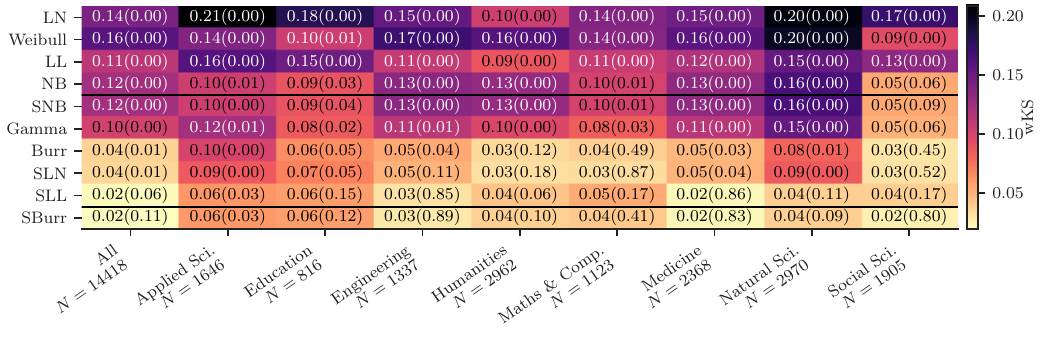}
    \caption[Distribution fit comparison]{\textbf{Distribution fit comparison.} We compare the optimized $\wks$-distance across common~\cite{kleiber.kotz_statisticalsizedistributions_2003} size distributions (rows) sorted and grouped by their parameter count. Abbreviations denote the log normal (LN), log logistic (LL), negative binomial (NB), shifted negative binomial (SNB), shifted log normal (SLN), shifted log logistic (SLL), and shifted Burr (SBurr) distributions. The $p$-values reported in parentheses show the fraction of simulated $\wks$-distances that are larger than the empirical $\wks$-distance when drawing and refitting $\NBS$ samples from the fitted distribution. High values ($p > 0.05$ marked by bold typeface) suggest that the empirical counts could be a sample from the fitted distribution. As expected, more complex distributions fit the data better. Several three parameter distributions, particularly $\SLN$ and $\SLL$, fit well across most domains. Only Applied Sciences is not well described by any of the tested distributions, even when the observed $\wks$-distances are relatively small.}
    \label{sfig:model_fit_heatmap}
\end{figure}

\begin{table}
    \centering
    \small
    \renewcommand{\arraystretch}{2.0}
    \setlength{\tabcolsep}{5pt}
    \begin{tabularx}{\linewidth}{l l >{\raggedright\arraybackslash}X}
        \toprule
        Distribution & PDF $q_s$ & Parameter bounds \\
        \midrule
        Weibull & $\dfrac{\beta}{\alpha} \left(\dfrac{s}{\alpha}\right)^{\beta-1} \exp\!\left(-\left(\dfrac{s}{\alpha}\right)^{\beta}\right)$ & $\alpha \in [0.01, 100]$, $\beta \in [0.01, 10]$ \\
        Negative binomial (NB) & $\dfrac{\Gamma(s+r)}{\Gamma(s+1)\,\Gamma(r)}\, p^{r} (1-p)^{s}$ & $r \in [0.5, 200]$, $p \in [0.01, 0.99]$ \\
        Log normal (LN) & $\dfrac{1}{s\,\sigma\sqrt{2\pi}} \exp\!\left(-\dfrac{(\ln s - \mu)^{2}}{2\sigma^{2}}\right)$ & $\mu \in [1, 50]$, $\sigma \in [0.1, 1]$ \\
        Log logistic (LL) & $\dfrac{\beta}{\alpha}\, \dfrac{(s/\alpha)^{\beta-1}}{\left[1 + (s/\alpha)^{\beta}\right]^{2}}$ & $\alpha \in [0.01, 100]$, $\beta \in [0.01, 7]$ \\
        Gamma & $\dfrac{(s-\ell)^{\alpha-1}}{\theta^{\alpha}\,\Gamma(\alpha)} \exp\!\left(-\dfrac{s-\ell}{\theta}\right)$ & $\alpha \in [0.01, 5]$, $\ell \in [0, 15]$, $\theta \in [0.01, 20]$ \\
        Shifted NB & $\mathrm{NB}(s + s_0;\, r, p)$ & $r \in [0.5, 200]$, $p \in [0.01, 0.99]$, $s_0 \in [0, 15]$ \\
        Shifted LN ($\SLN$) & $\mathrm{LN}(s + s_0;\, \mu, \sigma)$ & $s_0 \in [1, 30]$, $\mu \in [1, 50]$, $\sigma \in [0.1, 1]$ \\
        Shifted LL ($\SLL$) & $\mathrm{LL}(s + s_0;\, \alpha, \beta)$ & $\alpha \in [0.01, 100]$, $\beta \in [0.01, 9]$, $s_0 \in [1, 50]$ \\
        Burr & $\dfrac{c\,k}{\lambda}\, \dfrac{(s/\lambda)^{c-1}}{\left[1 + (s/\lambda)^{c}\right]^{k+1}}$ & $c \in [0.01, 10]$, $k \in [0.01, 10]$, $\lambda \in [0.01, 100]$ \\
        Shifted Burr & $\mathrm{Burr}(s + s_0;\, c, k, \lambda)$ & $c \in [0.01, 15]$, $k \in [0.01, 15]$, $\lambda \in [0.01, 100]$, $s_0 \in [1, 50]$ \\
        \bottomrule
    \end{tabularx}
    \caption[Candidate size distributions]{Candidate size distributions, their probability density $q_s$, and the parameter bounds used when fitting the empirical size distributions (see the main text). Each shifted variant reuses the corresponding unshifted density at $s + s_0$. Here $\Gamma$ denotes the gamma function, $\ell$ the gamma location, and $\lambda$ the Burr scale.}
    \label{stab:models}
\end{table}

\begin{table}
    \centering
    \begin{tabular}{llrrrrr}
\toprule
Domain & Model & $s_0$ & $\alpha$ & $\beta$ & $\mu$ & $\sigma$ \\
\midrule
All & SLL & 9.925 & 21.318 & 4.129 &  &  \\
Education & SLL & 13.303 & 23.484 & 5.802 &  &  \\
Applied Sciences & SLL & 18.137 & 28.273 & 6.795 &  &  \\
Humanities & SLN & 2.589 &  &  & 2.519 & 0.594 \\
Medicine and Health & SLL & 13.288 & 23.639 & 4.619 &  &  \\
Social Sciences & SLN & 6.930 &  &  & 2.965 & 0.432 \\
Engineering & SLL & 8.082 & 21.124 & 3.776 &  &  \\
Mathematics and Computing & SLN & 4.626 &  &  & 2.937 & 0.569 \\
Natural Sciences & SLL & 15.285 & 30.228 & 4.697 &  &  \\
\bottomrule
\end{tabular}

    \caption[Fitted parameters]{Fitted parameters for the $\SLL$ and $\SLN$ distributions across all departments and domains with uncertainty estimates in parentheses. The shift parameter $s_0$ controls the distribution location, while the shape parameters $\alpha$ and $\beta$ (for $\SLL$) or $\mu$ and $\sigma$ (for $\SLN$) capture the skewness and variability of the size distribution.}
    \label{stab:fit_params}
\end{table}

\clearpage
\starsection{Closure risks and growth rates}
\begin{figure}[h]
    \centering
    \includegraphics[width=0.6\textwidth]{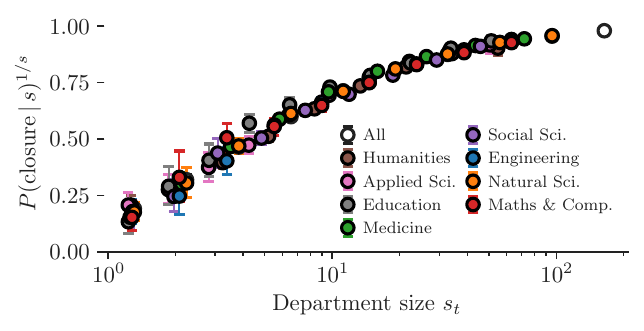}
    \caption[Closure risk expectation]{\textbf{Closure risk expectation.} Under an independent per-faculty annual attrition risk, we expect department closure to be more likely for smaller departments as it requires fewer individual attrition events. Under that assumption, $P(\mathrm{closure}\,|\,s)^{1/s}$ recovers the per-faculty attrition risk which would be constant in $s$ if closure were the result of individual attrition. The estimated per-faculty risk grows with size $s$ instead and it reaches unrealistically high values for large departments. Closure is not the result of individual attrition events, but rather size-dependent department-level events, such as mergers and splits. The implied per-faculty risk is identical across domains, suggesting that the size-dependence of closure risks is similar across scientific domains.}
    \label{sfig:closure_risk_expectation}
\end{figure}

\begin{figure}[h]
    \centering
    \includegraphics[width=\textwidth]{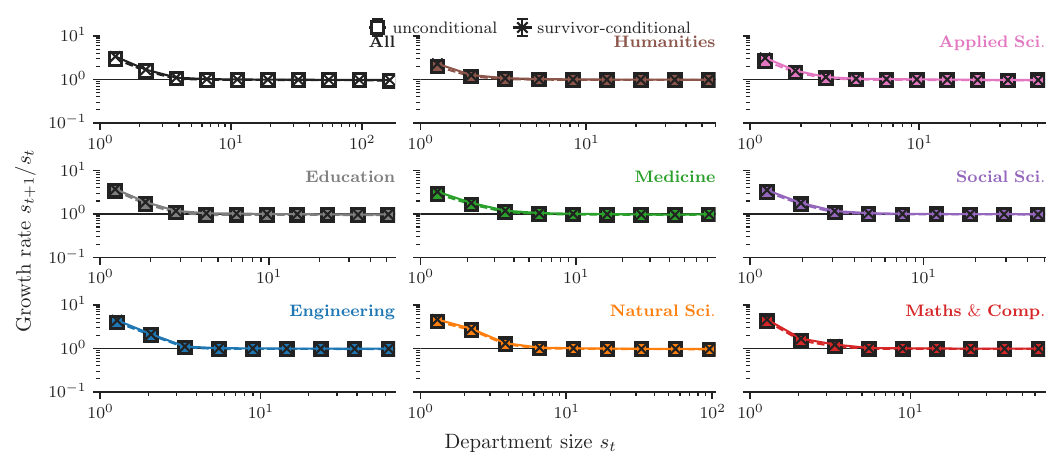}
    \caption[Growth rates conditional vs. unconditional]{\textbf{Growth rates conditional vs. unconditional.} Growth rate patterns are identical when conditioning on non-closure or not. This suggests that the observed growth patterns for small departments are not driven by closure events, but rather by other size-dependent factors.}
    \label{sfig:growth_rates}
\end{figure}

\end{document}